\documentclass[12pt,preprint]{aastex}

\usepackage{epstopdf}
\usepackage{graphicx}
\usepackage{natbib}
\usepackage{amsmath}
\usepackage{epsfig}
\usepackage{psfig}

\newcommand{\proper}[1]{{\it #1}} 
\newcommand{\cmd}[1]{{\tt #1}} 

\begin{document}

\title{{\it Chandra} detection of a new diffuse X-ray component from the globular cluster 47 Tucanae}

\author{
E.~M.~H.~Wu\altaffilmark{1},
C.~Y.~Hui\altaffilmark{2},
A.~K.~H.~Kong\altaffilmark{3},
P.~H.~T.~Tam\altaffilmark{3},
K.~S.~Cheng\altaffilmark{1},
V.~A.~Dogiel\altaffilmark{4}
}
\email{cyhui@cnu.ac.kr}
\altaffiltext{1}{Department of Physics, University of Hong Kong, Pokfulam Road, Hong Kong}
\altaffiltext{2}{Department of Astronomy and Space Science, Chungnam National University, Daejeon, Republic of Korea}
\altaffiltext{3}{Institute of Astronomy and Department of Physics, National Tsing Hua University, Hsinchu, Taiwan}
\altaffiltext{4}{I.~E.~Tamm Theoretical Physics Division of P.~N.~Lebedev Institute of Physics, Leninskii pr. 53, 119991 Moscow, Russia}

\begin{abstract}
In re-analyzing the archival \proper{Chandra} data of the globular cluster 47 Tucanae, we have detected a new diffuse X-ray emission 
feature within the half-mass radius of the cluster. The spectrum of the diffuse emission can be described by a power-law model 
plus a plasma component with photon index $\Gamma\sim1.0$ and plasma temperature $kT\sim0.2$~keV. While the thermal component 
is apparently uniform, the non-thermal contribution falls off exponentially from the core. The observed properties could possibly be 
explained in the context of multiple shocks resulted from the collisions among the stellar wind in the cluster and the 
inverse Compton scattering between the pulsar wind and the relic photons. 
\end{abstract}


\section{Introduction}
Globular clusters (GCs) are gravitationally bound dense systems hosting a very old stellar population of our Galaxy. 
Due to the high stellar densities in their cores, GCs are efficient in producing compact binaries through dynamical 
interactions (Pooley \& Hut 2006; Pooley et al. 2003; Hui et al. 2010). 
As millisecond pulsars (MSPs) are descendants of low-mass X-ray binaries (LMXBs) 
(Alpar et al. 1982), GCs are expected to be efficient factories of MSPs. 

In high energy regime, the Large Area Telescope (LAT) on board the \proper{Fermi} Gamma-ray Space Telescope has detected 
GeV $\gamma$-rays from a number of Galactic GCs (Abdo et al. 2009a, 2010; Kong et al. 2010; Tam et al. 2011), 
confirming GCs as a class of $\gamma$-ray emitters. There are two possible explanations for the 
origin of $\gamma$-ray emission from GCs. 
In one scenario, the emission can be interpreted as the collective magnetospheric emission from the whole MSP population 
in the GC (e.g. Venter \& de Jager 2008). Indeed, the $\gamma$-ray spectra of these detected GCs are reminiscent of those measured 
from the Galactic MSP population (e.g. Abdo et al. 2009b) as well as the individual MSPs detected in GCs NGC 6624 (Freire et al. 2001)
and M28 (Wu et al. 2013; Johnson et al. 2013). On the other hand, pulsar wind from the MSPs could also contribute 
to the observed $\gamma$-rays via the inverse Compton scattering (ICS) between relativistic pulsar wind leptons and background 
soft photons (Bednarek \& Sitarek 2007; Cheng et al. 2010; Hui et al. 2011).
The IC scattering between these wind particles and the relic photons can also result in a diffuse X-ray structure 
(Cheng et al. 2010). Searching for this predicted extended feature in X-ray can help to constrain the pulsar wind model of 
$\gamma$-ray GCs.

Hui et al. (2009) has searched for diffuse X-ray emission from the core regions of ten GCs using \proper{Chandra} archival data, 
enabling unresolved X-ray emission to be found for four clusters, which was attributed to unresolved faint point source populations 
in the cores of the clusters. 
On the other hand, Eger et al. (2010) have detected diffuse X-ray emission associated with Terzan 5 outside its 
half-mass radius. They found that the emission can be characterized by a power law with hard photon index 
$\Gamma \sim 1$ and is unlikely to be contributed from a population of unresolved point sources below the detection limit. 
The hard photon index was suggested to be a result of a non-thermal scenario by synchrotron radiation from energetics 
considerations. However, the limited statistics did not enable this study to draw a firm conclusion.

Motivated by the aforementioned discovery, Eger \& Domainko (2012) has systematically searched archival \proper{Chandra} data for 
diffuse X-ray emission from several Galactic GCs that has been detected in $\gamma$-rays by \proper{Fermi}. 
The authors excluded 47 Tuc in their analysis because less than half of the potential diffuse emission region 
(i.e. between one and three half-mass radius $r_{\text{hm}}$) is covered by the field of view (FoV) in the adopted datasets.

A previous search for diffuse X-ray emission in the direction of 47 Tuc has idenitified an extended feature beyond its half-mass radius
and in the direction of proper motion of the cluster (Okada et al. 2007). However, such feature has later been confirmed to be not 
physically associated with 47 Tuc (Yuasa et al. 2009). While these studies were focused on an individual extended feature at a 
location offset from the center of 47~Tuc, we are interested in searching for an isotropic diffuse X-ray component associated with 
this GC.

Being a nearby GC at 4.5 kpc (Harris 1996) and a host of a large MSP population, 47 Tuc is one of the promising 
targets to search for the diffuse X-ray feature as theoretically predicted (cf. Cheng et al. 2010). 
Together with the large volume of archival X-ray data, this motivates a deep search for diffuse emission in an unexplored 
region outside the core radius $r_{c}$ of the cluster.

In this Letter, we report the results from searching for diffuse X-ray emission associated with 47 Tuc in the region 
from $2r_{c}$ up to $\sim4\arcmin$ from the center. This selected field can alleviate the problem of potential contamination 
by the unresolved faint source population (cf. \S3) and allow a region sufficiently large for our search. 

\section{Data Analysis and Results}
To search for faint diffuse X-ray emission from 47 Tuc, we made use of archival \proper{Chandra} data and restricted 
our analysis on the Advanced CCD Imaging Spectrometer (ACIS) data with an exposure 
$>10$~ks. This limited us to four observations in 2002 (ObsID 2735, 2736, 2737 and 2738). The aim-points 
of all the four observations are on the S3 chip. We first reprocessed  
the data by using CIAO (Ver.~4.5) script \cmd{chandra\_repro} with the calibration files CALDB (Ver.~4.5.5). 
Inspections on the 0.5-7.0 keV lightcurves indicate that two observations (ObsID 2736 and 2738) suffered from prominent background 
flares. We subsequently performed the good time interval (GTI) filtering by using the CIAO routine \cmd{lc\_clean()}. 
Combining all the clean data resulted in an effective exposure of $\sim238$~ks.

By inspecting the reprocessed event files, signs of pileup effects were seen caused by the bright sources in 
the core of the cluster. We used the CIAO tool \cmd{pileup\_map} and the associated conversion formulae to deduce 
the pileup fractions.\footnote{See \url{http://cxc.harvard.edu/csc/memos/files/Davis\_pileup.pdf}} It was found 
that the three brightest sources detected by Heinke et al. (2005), X7, X9 and X5, have pileup fractions $> 5\%$, with 
that of X7 exceeding $10\%$. We excluded a rectangular region with a width of 0.5' centered on X7 and a length along 
the readout direction of the entire chip in subsequent spectrum extraction so as to avoid the contamination of the out-of-time 
events from the bright source.

\begin{figure}[b]
\plotone{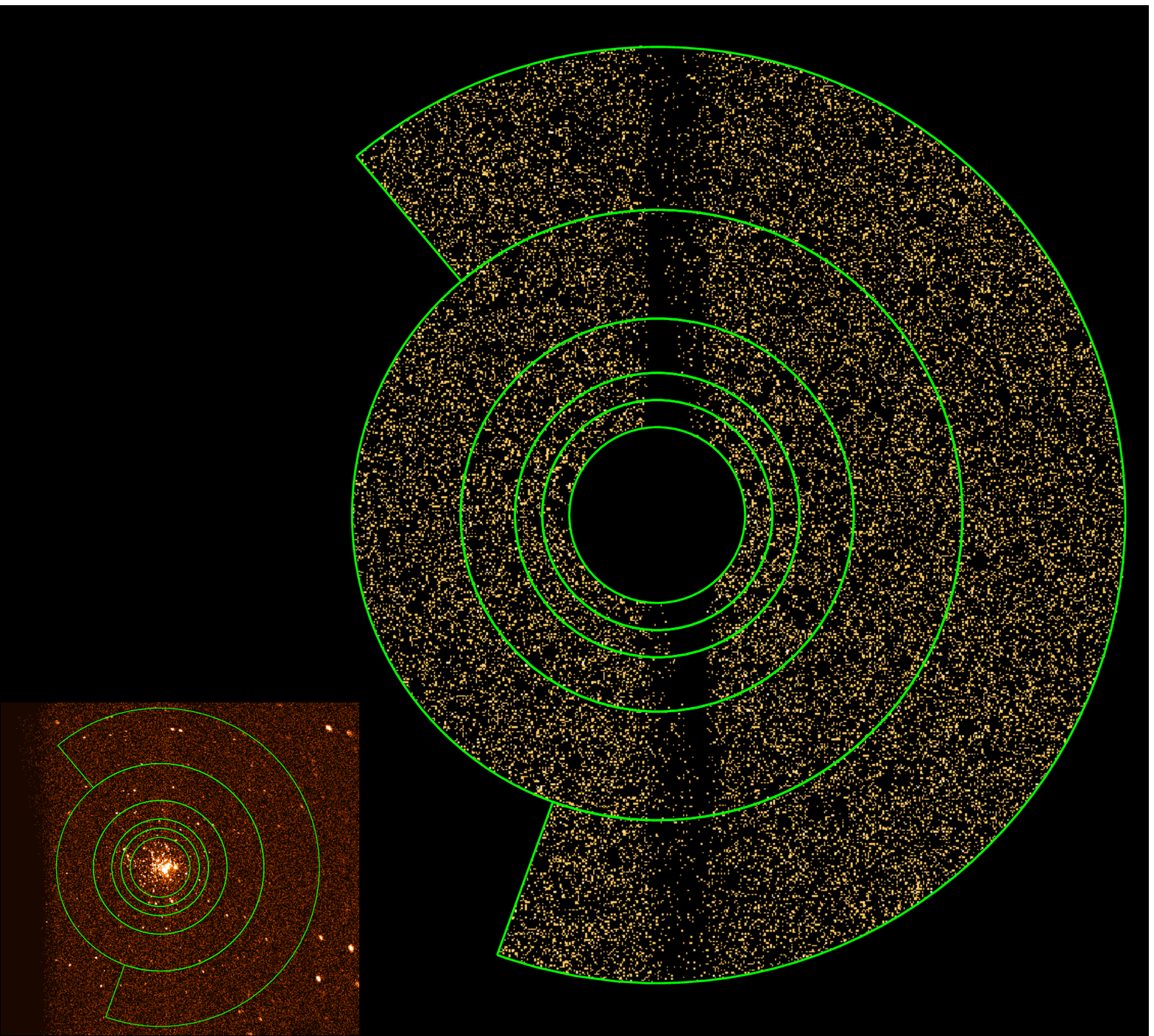}
\caption{Combined 0.5 -- 7.0 keV events of all four observations after point-source removal. 
The entire region-of-interest (RoI) is illustrated as an assembly of individual rings. 
The excluded rectangular regions for avoiding pileup contamination is not shown. 
The inset shows the combined event file in the same energy band before point-source removal as an illustration 
of the position of the RoI in the GC.}
\label{roirings}
\end{figure}

To define the region-of-interest (RoI), we first chose the largest annular region in $2r_{\text{c}} \leq r \leq 2\farcm5$ 
which is fully contained in the S3 chip. In order to fully utilize the FoV, we appended the aforementioned region with an 
truncated annular region extending to $3.75'$ (see Figure~\ref{roirings}). The angular span of this region was chosen 
such that it lies entirely within the CCD chip in all four observations. The entire RoI in our study was defined by assembling
all these adopted regions.

To effectively remove the resolved point 
sources in the RoI, we took the sky positions in the point source catalog by Heinke et al. (2005) and calculated 
the size of the \proper{Chandra} point spread function (PSF) at 1.5 keV with an enclosed-counts fraction (ECF) 
of $96\%$ at the positions of these point sources by using 
the \cmd{psf} \proper{Python} module. Since Heinke et al. (2005) covers only sources 
within $2\farcm79$ of 47 Tuc, we ran \cmd{wavdetect} in the energy band 0.5-7.0 keV to detect sources beyond 
this region. These sources were subsequently removed as aforementioned. 

After removing all the resolved point sources and excluding the pileup-affected region, we extracted the spectra from 
individual observations within the entire RoI in an energy band of $0.5-7.0$~keV. 

We began our investigation with a spectral analysis of the X-ray emission from the entire RoI. 
For tightly constraining the emission properties, we fitted the spectra extracted from all observations simultaneously. 
Since the RoI covers a large fraction of the S3 chip, there are limited source-free regions on the chip for extracting the 
background spectra. In addition, the peripheral region of the chip could possibly 
contain the contribution of diffuse X-ray emission that we are searching for. 
Therefore, we adopted the ACIS blank-sky event files as the background throughout our study. 
We followed the online analysis thread to tailor and reproject the blank-sky background files so as to match the observed 
datasets\footnote{http://cxc.harvard.edu/ciao/threads/acisbackground/}. 
 
For judging the background normalization, we have inspected the source and background spectra individually for each observation 
in $10-12$~keV. In this band, observed data are dominated by particle background. Except for the data from 
ObsID~2738, the estimated background contributions for all other observations are $>99.7\%$ which indicates the particle 
background normalizations are legitimate. For ObsID~2738, we noticed a residual background ($\gtrsim6\%$) in this band. 
For minimizing the systematic uncertainties, we excluded this data in all subsequent analysis.
The response files used in this extended 
source analysis were commputed by weighting the flux contribution from each pixel with the aid of a detector map.

\begin{figure}[t]
\includegraphics[angle=270,width=\textwidth]{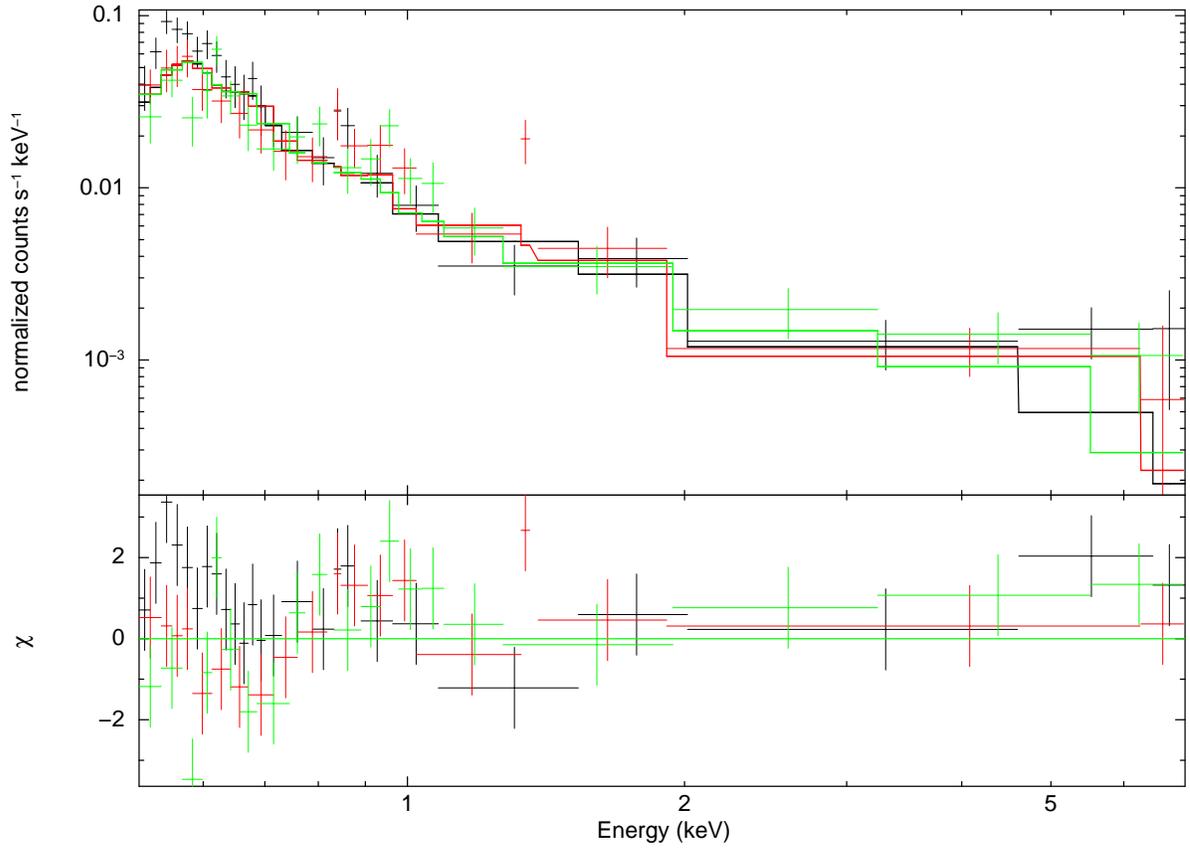}
\caption{({\it Upper panel}): Spectrum of the unresolved X-rays in the region-of-interest, 
fitted by an absorbed power-law model plus a MEKAL plasma component. ({\it Lower panel}): Contributions to 
the $\chi^{2}$ fit statistic are shown. }
\label{roispec}
\end{figure}

For all the spectral analysis, we fixed the column absorption at 
$N_{\mathrm{H}} = 2.2 \times 10^{20}~\mathrm{cm}^{-2}$ which was deduced from the foreground reddening 
$E(\bv) = 0.04$ (Harris 1996; Predehl \& Schmitt 1995; Cardelli et al. 1989). 
We first fitted the spectra using an absorbed power law (PL) model in a form of $KE^{-\Gamma}$, where the free 
parameters $K$ and $\Gamma$ are the normalization in unit of photons~keV$^{-1}$~cm$^{-2}$~s$^{-1}$ at 1~keV and the index 
of the photon flux respectively. The best-fit yields a photon index of 
$\Gamma = 4.48^{+0.17}_{-0.16}$. However, we noticed that the goodness-of-fit is not desirable.
We have also attempted to fit the data with a thermal plasma model which includes the contribution from the metal lines 
(XSPEC model: MEKAL). 
But we found this model also results in a poor goodness-of-fit. 
On the other hand, by adding a MEKAL component to the PL model, a reasonable description of the observed data can be 
obtained ($\chi^{2}_{\nu}=1.15$ for 415 d.o.f.). This composite model yields a photon index 
of $\Gamma = 0.96^{+0.34}_{-0.29}$ and a plasma temperature 
of $kT = 0.19 \pm 0.01$. $F$-test indicates that the PL+MEKAL model is favored over a single PL or a 
single MEKAL model at a confidence level $>7\sigma$. The energy spectrum of the entire RoI with the PL+MEKAL model fit 
is shown in Figure~\ref{roispec}.

In order to estimate the contribution from unresolved point sources in the RoI, we constructed the cumulative luminosity 
function by using the luminosities of the sources within $r_{c}$ in 0.5-6~keV reported by Heinke et al. (2005).
Fitting the distribution with a functional form of $N(>L_{X})=N_{0}L_{X}^{\alpha}$, we obtained 
$\alpha=-0.55\pm0.01$ and $\log N_{0}=18.26\pm0.19$. 
With a putative detection limit 
of $8\times10^{29}$ erg s$^{-1}$ (Heinke et al. 2005), we integrated the source density in the 0.5 -- 6.0 keV energy range to 
obtain a luminosity of $1\times10^{32}$ erg s$^{-1}$ within $r_{\text{c}}$. 
For estimating their contribution in our RoI, we adopted a radial source density with a form of:
\begin{equation}
  s(r) = s_0 \left[1 + \left(\frac{r}{r_{\text{c}}}\right)^2\right]^{(1 - 3q)/2}~,
\end{equation}
\noindent where $s_0$ is a normalization factor and $q = 1.63$ is the best-fit value
obtained by Heinke et al. (2005). 
The expected luminosity of the unresolved point sources can then be estimated by 
$10^{32}~\left(\int_{2r_{\text{c}}}^{3\farcm835}{s(r)\,\mathrm{d}r} / \int_{0}^{r_{\text{c}}}{s(r)\,\mathrm{d}r}\right)$ erg s$^{-1} \approx 10^{31}$ erg s$^{-1}$. To compare with this value, we calculated the measured luminosity in the same energy range 
to be $\left(3.82^{+0.31}_{-0.30}\right)\times10^{32}$ erg s$^{-1}$ for the PL+MEKAL model. Therefore, we conclude that 
the contribution from unresolved point sources to the observed emission is negligible. 

\begin{figure}
\vspace{-7cm}
\begin{center}
\epsfig{figure=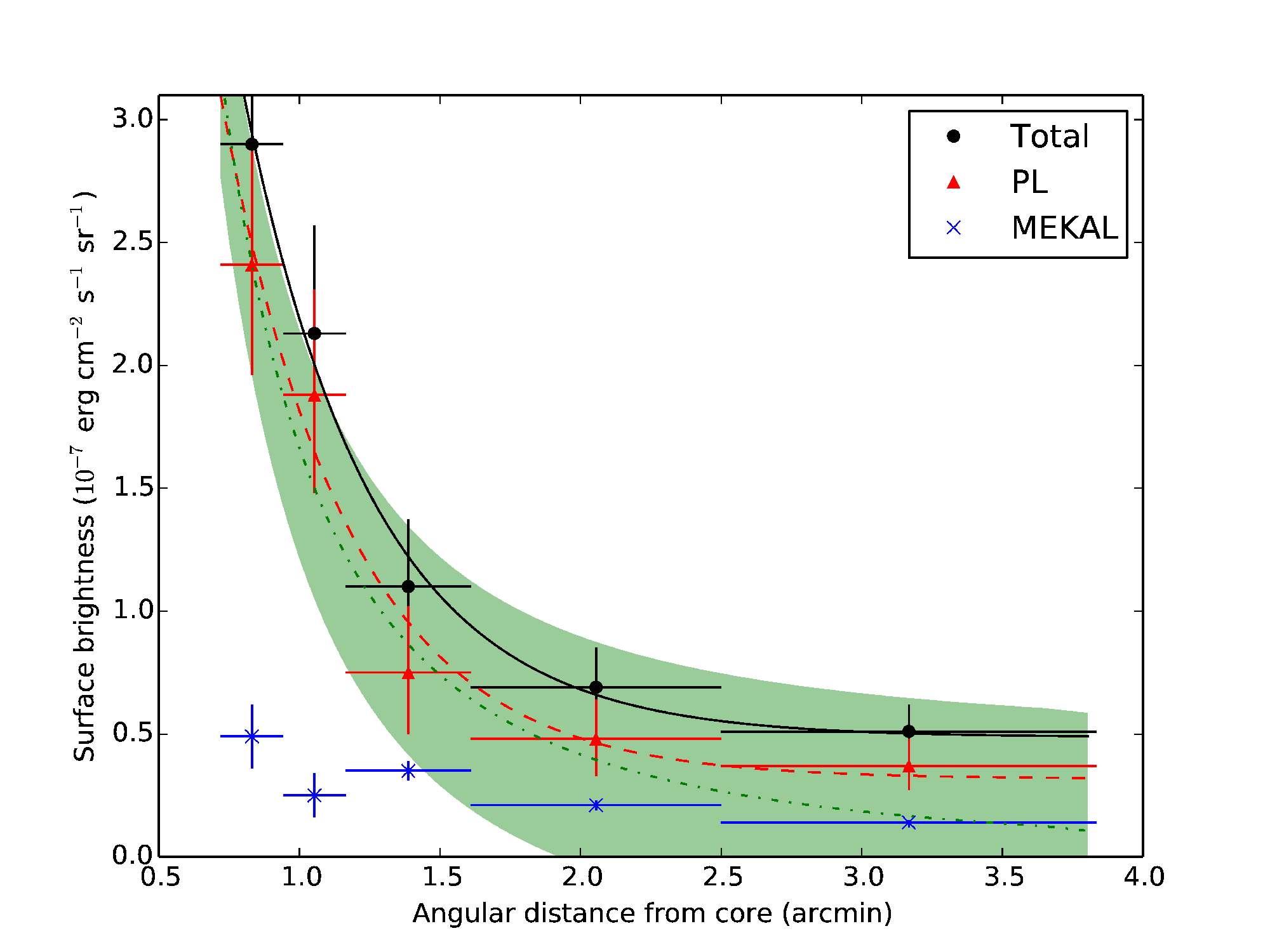,width=15cm}
\caption{\footnotesize Unabsorbed diffuse X-ray surface brightness profile in the 0.5 -- 7.0 keV band (black circles). Red triangles and blue crosses represent the brightness of the PL component and the MEKAL component, respectively. The black solid line and the red dashed line show, respectively, the best-fit exponential profiles of the total brightness and the PL component. The green dot-dashed line denotes the stellar density profile from Michie (1963) and Kuranov \& Postnov (2006). 
It is scaled to match the first data point of the PL component and the uncertainty 
of this profile due to the error of the matched data point is shown by the green-shaded region.}
\end{center}
\label{radprofile}
\end{figure}

To quantify the distribution of the diffuse X-ray emission as an angular distance from the center of the cluster, 
we divided the RoI into five concentric annular regions (rings). The sizes of each ring were chosen to obtain the spectral 
parameters with comparable relative errors in each ring. 
The fitting results of all rings are summarized in 
Table~\ref{fitrings}. To obtain a surface brightness profile, we integrated the best-fit models for each ring 
over the energy band 0.5 -- 7.0 keV and divided the unabsorbed flux by the corresponding solid angle of the 
extraction region. The solid angles were computed with the exclusion of the PSF subtraction regions of point sources and 
the bad pixels. 
The brightness profile is plotted in Figure~\ref{radprofile} in which the total surface brightness has shown 
an exponential fall-off. For a more detailed investigation, we disentangled the thermal and non-thermal contributions as 
shown in Fig.~\ref{radprofile}. 
While the non-thermal X-rays follows an exponential fall-off that conforms with 
the stellar density profile of 47~Tuc (green dot-dashed line in Fig.~\ref{radprofile}), the plasma component is 
apparently uniform.

To quantify the structure of the diffuse emission, we have fitted the function 
$S(r) = S_0 \exp{(-r/r_{\text{falloff}})} + C$ to the brightness profile, 
where $S_0$ is a normalization parameter, $r_{\text{falloff}}$ is the characteristic fall-off distance, and $C$ is the brightness 
at the tail. The best-fit for the total brightness yields $r_{\text{falloff}} = 0\farcm46 \pm 0\farcm08$ and that for the PL component gives $r_{\text{falloff}} = 0\farcm45 \pm 0\farcm15$, which are compatible within uncertainties. On the other hand, the 
best-fit constant levels for the total brightness and the PL component are found to be 
$C=(4.9\pm1.2)\times10^{-8}$~erg~cm$^{-2}$~s$^{-1}$~sr$^{-1}$ and $C=(3.2\pm2.1)\times10^{-8}$~erg~cm$^{-2}$~s$^{-1}$~sr$^{-1}$ 
respectively. 

\begin{deluxetable}{ccccccc}
\tablewidth{0pc}
\tablecaption{Spectral fitting results of the entire RoI.}
\startdata
\hline\hline
Model & $\chi_\nu^2$ (d.o.f.) & Norm & $\Gamma$ & $kT$ & $L_{\rm PL}$ & $L_{\rm MEKAL}$ \\
 & & & & (keV) & ($10^{32}$ erg s$^{-1}$) & ($10^{32}$ erg s$^{-1}$) \\
\hline
PL       & 1.25 (417) & $1.95^{+0.13}_{-0.13}$ & $4.48^{+0.17}_{-0.16}$ & \nodata                & $1.70^{+0.06}_{-0.07}$ & \nodata                \\
MEKAL    & 1.37 (417) & $4.49^{+0.26}_{-0.26}$ & \nodata                & $0.20^{+0.01}_{-0.01}$ & \nodata                & $1.43^{+0.06}_{-0.06}$ \\
PL+MEKAL & 1.15 (415) & $0.90^{+0.20}_{-0.17}$ / $4.38^{+0.27}_{-0.27}$ & $0.96^{+0.34}_{-0.29}$ & $0.19^{+0.01}_{-0.01}$ & $2.36^{+0.50}_{-0.48}$ & $1.26^{+0.07}_{-0.08}$ \\
\enddata
\tablenotetext{a}{\footnotesize{Normalization parameter of the PL/MEKAL model/component. It is $10^{-5}$ photons cm$^{-2}$ s$^{-1}$ keV$^{-1}$ at 1 keV for a PL model. It is equal to $(10^{-14} / 4\pi D^2) \int{n_{\text{e}} n_{\text{H}}} \mathrm{d}V$ for a MEKAL model, where $D
$ is taken to be 4.5 kpc.}}
\tablenotetext{b}{\footnotesize{Plasma temperature.}}
\tablenotetext{c}{\footnotesize{Integrated luminosity of the PL component in the energy 0.5 -- 7.0 keV.}}
\tablenotetext{d}{\footnotesize{Same as (c) but for the MEKAL component.}}
\label{fitresults}
\end{deluxetable}

\begin{deluxetable}{cccccccc}
\tablewidth{0pc}
\tablecaption{Spectral fitting results of individual rings by using PL+MEKAL model.}
\startdata
\hline\hline
Ring & $r_{\text{min}}$\tablenotemark{a} & $r_{\text{max}}$\tablenotemark{b} & $kT$ (keV) & $\Gamma$ & $S_{\text{PL}}$\tablenotemark{c} & $S_{\text{MEKAL}}$\tablenotemark{d} & $S_{\text{total}}$\tablenotemark{e} \\
\hline
1 & 0.72   & 0.9425 & $0.21\pm0.03$  & $0.77^{+0.33}_{-0.32}$ & $2.41^{+0.48}_{-0.45}$ & $0.49\pm0.13$ & $2.90^{+0.50}_{-0.47}$ \\
2 & 0.9425 & 1.165  & $0.17^{+0.04}_{-0.07}$  & $0.90^{+0.33}_{-0.32}$ & $1.88^{+0.43}_{-0.40}$ & $0.25\pm0.09$ & $2.13^{+0.44}_{-0.41}$ \\
3 & 1.165  & 1.61   & $0.25\pm0.02$  & $0.47^{+0.62}_{-0.61}$ & $0.75^{+0.27}_{-0.25}$ & $0.35\pm0.04$ & $1.10^{+0.27}_{-0.25}$ \\
4 & 1.61   & 2.5    & $0.18\pm0.01$  & $0.36^{+0.54}_{-0.55}$ & $0.48^{+0.16}_{-0.15}$ & $0.21\pm0.02$ & $0.69^{+0.16}_{-0.15}$ \\
5 & 2.5    & 3.835  & $0.18^{+0.01}_{-0.02}$  & $0.85^{+0.34}_{-0.40}$ & $0.37^{+0.11}_{-0.10}$ & $0.14^{+0.01}_{-0.02}$ & $0.51^{+0.11}_{-0.10}$ \\
\enddata
\label{fitrings}
\tablenotetext{a}{Inner radius of annular region in arcminutes.}
\tablenotetext{b}{Outer radius of annular region in arcminutes.}
\tablenotetext{c}{Surface brightness for the PL component in $10^{-7}$~erg~cm$^{-2}$~s$^{-1}$~sr$^{-1}$ over the energy range of 0.5 -- 7.0 keV.}
\tablenotetext{d}{Same as (c) but for the MEKAL component.}
\tablenotetext{e}{Total surface brightness as a sum of (c) and (d).}
\end{deluxetable}

\section{Summary \& Discussion}

In examining the archival \emph{Chandra} data, we have uncovered a faint extended X-ray feature in a region from $2r_{c}$ to $\sim4'$ 
from the center of 47~Tuc. 
Figure 3 shows that the unabsorbed diffuse X-rays in the 0.5-7.0 keV band consist of two components, i.e. an exponential fall-off 
non-thermal component plus a uniform thermal component. 
The spectrum of the non-thermal component is very hard with a photon index $\Gamma \sim 1$. 
The surface brightness of this component apparently correlates wtih the point source distribution (i.e. Equation~1) very well. 
The profiles of both extended feature and the radio point source density fall by $\sim90\%$ from 1' to 2' and $\sim98\%$ from 1' 
to 3'.

It is known that stellar winds of massive stars are clumpy and non-thermal emission has been observed 
(cf. Feldmeier 2001 for a general review). In the core of a GC, collisions among stellar winds from high 
stellar density region could create multiple shocks that resemble the situation of stellar wind from massive stars. 
The electron distribution in multi-shock regions is known to have a very hard distribution 
($\frac{dN_{e}}{dE_e} \sim E_e^{-1}$) (Bykov \& Fleishman 1992; Bykov \& Toptygin 1993). 
For a power-law distribution of electrons: $\frac{dN_{e}}{dE_e}\propto E_{e}^{-\alpha}$, the radiation via either synchrotron 
radiation or inverse Compton scattering will have a photon index of $\Gamma=\frac{\alpha+1}{2}$. 
And hence, a very hard X-ray spectrum is expected from the multi-shock regions. 

To estimate the shock radius $R_{\rm sh}$, we used Equation~(5) of Weaver et al. (1977):
$R_{\rm sh}\sim(L_{w}\tau^{3}/\rho_{\rm ISM})^{1/5}$, 
where $L_{w}$ and $\tau$ are the total wind power from the cluster and the expansion time scale respectively. 
The wind power depends on the number of stars in the GC $N_{*}$, the mass-loss rate $\dot{M}$ and 
the wind velocity $v_{w}$: $L_{w}\sim N_{*}\dot{M}v_{w}^{2}/2$. 
Taking the characteristic values of stellar wind, i.e. $\dot{M}\sim10^{-14}M_{\odot}$~yr$^{-1}$ and $v_{w}\sim500$~km/s, which gives
$L_{w}\sim10^{33}(N_{*}/10^{6})$~erg/s. 47~Tuc has a finite proper motion at the order of $v_{p}\sim150$~km/s (cf. Okada et al. 2007). 
Assuming the shock maintains 
a spherical symmetry, the time scale can be estimated by Equation~(64) of Weaver et al. (1977) which gives
$\tau\sim10^{12}(n_{\rm ISM}/10^{-3}{\rm cm}^{-3})^{-1/2}(L_{w}/10^{33}~{\rm erg/s})^{1/2}(v_{p}/100~{\rm km/s})^{-5/2}$~s, 
where $n_{\rm ISM}\sim\rho_{\rm ISM}/m_{p}$ and $m_{p}$ is the proton mass. 
Substituting all these into $R_{\rm sh}$, the shock radius is found at an order of $\sim10$~pc. At 4.5~kpc, this corresponds 
to an angular size of $\sim8'$. Therefore  the total stellar wind coming out the core of the GC can form shock with ISM at a distance 
much larger than $r_{c}$. 

Consequently the radiation away from the core should consist of two components, i.e. the non-thermal component 
emitted from non-thermal electrons accelerated in the diffuse shock region, whose energy index is 2-3 (Bell 1978), and the thermal 
component emitted from the shock-heated plasma. 
The temperature behind shock is estimated by $T\sim3\mu V_{\rm sh}^{2}/16k$ for adiabatic shock (Weaver et al. 1977), 
where the shock velocity $V_{\rm sh}$ is estimated by $V_{\rm sh}\sim dR_{\rm sh}/dt \sim 3R_{\rm sh}/5\tau\sim3\times10^{7}$~cm/s. 
And hence $T\sim2\times10^{6}$~K which is consistent with the observed value. We can also estimate the thermal X-ray luminosity 
from thermal bremsstralung which is a function of local plasma density, plasma temperature and the total volume-of-interest 
(Rybicki \& Lightman 1979): $L_{x}\sim10^{-27} n^{2} T^{1/2} V$. For estimating the plasma density, 
we can utilize the normalization parameter deduced from the best-fit plasma model 
$\int n_en_H dV \sim 10^{55} {\rm cm}^{-3}$. Assuming 
$n_e\sim n_H$, we obtain $n_e \sim 0.03 (V/V_{\rm RoI})^{-1/2}$~cm$^{-3}$, where we adopted the radius of the RoI as
$r_{\rm RoI}\sim 5$~pc. We noticed that this estimate is less than the electron density ($n_{e}\sim0.07$~cm$^{-3}$) 
obtained by Freire et al. (2001). The under-estimation can stem from ignoring the metal content in the cluster. Adopting 
$T\sim2\times10^{6}$~K, $n\sim0.07$~cm$^{-3}$ and $V\sim2\times10^{58}$~cm$^{3}$, the thermal bremsstralung luminosity is at the order 
of $L_{x}\sim10^{32}$~erg/s which agrees with the observed value. 


Apart from the shock produced by the stellar wind and ISM, we have also considered other 
possible contribution to the non-thermal component. Cheng et al. (2010) has shown that 
inverse Compton (IC) scattering between the background soft photon 
fields and the relativistic electrons in pulsar wind can contribute to the 
$\gamma$-ray from GCs. They predict that a uniform X-ray component should exist due to 
IC scattering between the relic photons and the pulsar wind relativistic electrons. And the X-ray flux 
is proportional to the $\gamma$-ray flux as  
$F_x(5~{\rm keV}) \approx (5~{\rm keV}/8\gamma_{w5}^2~{\rm MeV})^{0.5} (w_{relic}/w_{soft})F_{\gamma}^{obs}$, 
where $\gamma_{w5}$ is the Lorentz factor of pulsar wind in unit of $10^{5}$, 
$F_{\gamma}^{obs}$ is the observed $\gamma$-ray energy flux in GeV regime, 
$ L_{34}$ is the typical spin-down power of MSP in units of $10^{34}$erg/s and 
$f_{e^{\pm}} \sim 30$ is the energy ratio between electrons and protons. 
Lorentz factor of pulsar wind is estimated as $\gamma_w = 2\times 10^5 f_{e^{\pm}}^{-1} L_{34}^{1/2} \sim 6\times 10^3$. 
Taking the ratio between the relic photon energy density $w_{relic}$ and soft photon energy density $w_{soft}$ 
to be $(w_{relic}/w_{soft})\sim 1/2$ (Cheng et al. 2010) and 
$F_{\gamma}^{obs} \sim 10^{-11}$ erg~cm$^{-2}$~s$^{-1}$ (Abdo et al. 2010), we obtain 
$F_x(5~{\rm keV}) \sim 3\times 10^{-12}$ erg~cm$^{-2}$~s$^{-1}$. Since $\gamma$-ray emission 
from 47~Tuc is consistent with a point source as observed by {\it Fermi}, hence its emission radius must be less than 
the PSF of the telescope (i.e. $\lesssim30'$). 
The corresponding IC X-ray luminosity emitted from $5'$ is given by 
$L_x(r<5')\sim F_x 4\pi D^2(5/30)^2\sim 3\times 10^{32}$erg/s, where $D\sim 5$~kpc is the distance to 47~Tuc, 
which is consistent with the observed non-thermal luminosity given in Table~\ref{fitresults}. 

Follow-up investigations of the nature of the extended X-ray feature of 47~Tuc are encouraged. 
Weaver et al. (1977) suggested that 
a large number of emission lines should be emitted from the shock-heated plasma. Deep X-ray observation with high spectral resolution 
is required to resolve them. 
Based on the IC model, radio flux from the pulsar wind should be also correlated with $\gamma$-ray flux (Cheng et al. 2010), future 
radio observations with higher resolution and sensitivity can also help to constrain the model. 

\acknowledgments{
CYH is supported by the National Research Foundation of Korea through grant 2011-0023383. 
AKHK is supported by the National Science Council of the
Republic of China (Taiwan) through grant NSC100-2628-M-007-002-MY3 and
NSC100-2923-M-007-001-MY3.
PHT is supported by the National Science Council of the Republic of China (Taiwan) through grant NSC101-2112-M-007-022-MY3.
KSC are supported by a GRF grant of HK Government under HKU7009 11P.
VAD acknowledges support from the RFFI grant 12-02-00005. 
}

\end{document}